# The pernicious danger of cortical brain maps

Benjamin Yost Hayden

Department of Neuroscience,
Center for Magnetic Resonance Research, and
Department of Biomedical Engineering
University of Minnesota, Minneapolis MN 55455

**Corresponding author:**
    Benjamin Yost Hayden
    Department of Neuroscience
    University of Minnesota, Minneapolis MN 55455
    Email: benhayden@gmail.com

**Competing interests**
    The author has no competing interests to declare.

**Acknowledgements**
    This research was supported by NIH grant R01 MH129439 and MH125377.

*"And then finally I did a remarkable, intuitive thing, which is I took the music I had written down and I erased all the bar lines. And suddenly, I saw something which I hadn't seen before, which was that I saw the patterns. It went over bar lines." –*

- Philip Glass

**The parcellation of the cerebral cortex**

The parcellation of the primate cerebral cortex into numbered regions, based on cytoarchitecture, began with the pioneering research of neuroanatomist Kobrinian Brodmann (Brodmann, 1909; Loukas et al., 2011). While the borders between regions have changed somewhat, and in some cases continue to be disputed, the idea of dividing the cortex into distinct numbered areas has become central to the goal of understanding brain function (Posner et al., 1998). And indeed, areal maps do provide a good starting point for functional parcellation. It is obvious, for example, that visual area V1 has a different function than primary motor cortex. However, as with anything good, one can take things too far (Wilson et al., 2010). Indeed, cortical areas, while useful, have several pernicious side effects for neuroscientists interested in function, especially in prefrontal cortex.

In short, areal maps ground our thoughts about brain function and short-circuit our imaginations. We see the borders drawn on the cortical surface and assume they are the most important organizing principle of cortical function, and become focused on telling a story in which those borders are the main characters. Instead, area borders are just one of several ways in which the cerebral cortex is organized. Neuroscientists now understand – much more so than we did at the time of Brodmann – that even if brain areal maps are *true*, they are far from the whole truth.

**Why are brain areas problematic?**

First, they are too large. What appear, on a map, to be single brain regions each contains multiple subareal organizational features. These include patches within the region with different functions, functions specific to certain layers, and functions specific to certain cell types. To give one example, our lab showed that the central part of the orbitofrontal cortex (OFC) can be subdivided into subregions that do not correspond to traditional cytoarchitectonic borders, based on connectivity, and that these subregions have important functional differences (Wang et al., 2022). These differences are as large as – or larger than – the differences between adjacent regions. To give another example, an important and influential discovery comes from face patches, which are important and different from surrounding tissue, but are smaller than traditionally recognized brain areas (Tsao et al., 2008; Grimaldi et al., 2016; Arcaro et al., 2020). These examples are just a few of the undoubtedly large set of subareal functional features that do not appear on most border maps.

Second, if they are too large, brain areas are also too *small*. The parcellation of the brain into areas discourages us from recognizing larger cross-area functional domains, including groups of adjacent or connected but non-adjacent regions with overlapping functions (Averbeck et al, 2014; Wilson et al., 2010; Graziano et al, 2002; Kim and Shadlen, 1999; Chafee et al., 1998). Sometimes that means there is a smooth continuum of function and organization at the macro-level. One well-known example comes from the dorsoventral functional gradients observed within the prefrontal cortex. There appear to be gradual – but not categorical - changes in information as it moves from ventral surface to the dorsal one, changes that are quantitative, not qualitative (Maisson et al., 2021; Fuster, 2000 and 2001). Another example comes from the ventral ("what") pathway of the visual cortex. While early studies of the visual cortex emphasized the distinct roles of its areas in qualitatively distinct gestalt features, more contemporary studies emphasize the gradual transformation of information along the ventral pathway (Van Essen, 1983; DiCarlo et al., 2012). Border maps obscure these macro-areal organizational features.

These issues are if anything growing with new methods of recording. Technologies that allow for recordings across hundreds of neurons have so far demonstrated largely preserved functions across multiple regions (e.g. Musall et al., 2019; Steinmetz et al., 2019; Stringer et al., 2019; Ottenheimer et al., 2002). And indeed, the "true" functions of these regions come from small residual effects after accounting for the much larger effects that are shared. From the classic perspective of brain areas, these results are puzzling – why wouldn't we see qualitatively different functional patterns in what are ostensibly different regions?

Even when looking at the level of organization beyond the brain area, it would be a mistake to limit ourselves to gradients, which are agglomerations of adjacent areas. Supra-areal functional structures do not necessarily have to be connected (Pessoa, 2014). It is well established that the brain contains specific functionally coherent networks of areas that are not adjacent on the map, but that may be supported by long-range highly specific anatomical connections. These networks, such as the default mode network, the attentional orienting network, and the salience network, are invisible on areal maps, and may therefore be deprioritized by researchers.

**Maps are usually based on cytoarchitecture not connectivity**

These limitations of areal maps are reflective of the core underlying weakness of areal maps, which is that they are based on cytoarchitecture, which has only a limited relationship with function. Instead, function is determined largely by connectivity. A neuron's response repertoire is, to oversimplify a bit, a weighted sum of its inputs. Unfortunately, not enough information about connectivity is known to make coherent areal maps of the brain. Indeed, this lack of knowledge is one reason for the continued dominance of cytoarchitectural maps. However, based on what we do know, there is likely to be a lot of difference between connectivity-based maps and cytoarchitectural maps.

Despite this limitation, areal maps encourage us to approach neuroscience from the perspective of the brain area, rather than brain function. That is, when we see a brain map, we naturally assume that each area has some specific function, and our job is to identify that function. This interest may be tempting even if we know that a one-to-one correspondence between areas and functions is correct. If some functions are distributed and some are specific to regions, we become more interested in the functions that correspond to regions, even if those functions are less important, in the grand scheme of things. In other words, areal maps encourage us to take a *structure-first approach* to understanding neural function. The alternative is a *function-first approach*. If we were to start with a specific brain function and ask how it is implemented, we would no doubt consider that it is implemented by a specific area, but would also consider other possibilities, such as broad distribution.

Indeed, it often seems as if neuroscientists' only goal is to identify the canonical (or even unique) function of each area. However, this approach is necessarily limited. A brain region may (and likely does) participate in several functions, many of which are shared broadly with others (for more, see Uttal, 2001). If we limit ourselves to functions uniquely performed by each area, we may inadvertently ignore almost all of its repertoire. Indeed, this approach encourages us to focus on whatever brain functions are narrowly realized, rather than ones that are more distributed – not because they are intrinsically more important, but because they fit our notions of area to function correspondence.

To give an example, in economic choice, early theories tried to fit different presumed economic functions, such as evaluation, comparison, selection, and monitoring to specific regions (Rangel et al., 2008). It has become abundantly clear since then that most prefrontal regions participate in all of these functions (Yoo and Hayden, 2018; Fine and Hayden, 2021). These findings likely indicate that choice is the result of distributed computations across many areas. The implication **should be** that we look for the neural mechanisms for how evaluation and so on take place. Instead, the field has become more focused on delineating subtle differences between different regions. These differences may be true, but they are missing the forest for the trees – the choice mechanism is ignored because it does not correspond cleanly to area.

**How strong is the evidence that brain areas are meaningful?**

These facts point to the danger posed by overreliance on areal maps in leading to a troubling form of circularity. The reliance on areal maps lead to research practices that seek to reify them, and in doing so, serve to reinforce their importance and centrality, leading to even greater confidence in their reality and important. This occurs because much research practice is not generally devoted to testing the *arealization hypothesis* (the idea that areal borders are the most important functional organization feature), but instead takes it as a given and delineates its properties. And by doing so they wind up strengthening the case for areas' borders.

Let us consider a well-known example to see how research practice reifies the area-function linkage: conflict monitoring and resolution. These two closely related processes have long been associated with the dorsal anterior cingulate cortex (Ebitz and Hayden, 2016), even though these processes are likely mediated by a wider swathe of areas, of which dACC may not have a particularly special role. And, indeed, it is clear that dACC has a much larger repertoire, of which its contribution to conflict processing is a small part (Heilbronner and Hayden, 2016). But the linkage between dACC and conflict is so close in the literature, if not in the brain. that scholars researching conflict are generally inclined to focus on the dACC. This is a reasonable research decision because dACC *does* show conflict effects (at least in its hemodynamic responses). But the result is a vicious cycle where the bulk of research on the area is devoted primarily to conflict. As researchers interested in conflict focus on dACC, their discoveries open up new questions about the nature of conflict and the role of the dACC in that process. Researchers interested in other topics will avoid the dACC, because they don't want to have to argue that their effects are confounded with conflict. And as time continues, each area of cortex becomes associated with one element of cognition. As a result we wind up with a functional map of cortex that makes it look more modular than it really is - the modularity reflects the structure of how the scientific enterprise is organized, not the organizational structure of the brain.

### How to do good neuroscience with less reliance on brain areas

None of this is to say that we should throw out brain area maps. We don't want to throw out the baby with the bathwater, and brain maps do have a lot to offer. But we need to understand their limitations, and we need to be willing to ignore them sometimes or often. But how do we actually do that?

We should begin by de-emphasizing the goal of assigning functions to area. We should emphasize the goal of delineating basic principles by which coordinated groups of neurons can solve important problems. We can take as a tonic example the case of attention. The field of attention arose after, and in parallel with, the development of the field of vision. The field began as an attempt to identify the site at which attention modulated neural activity. In other words, attention research has been successful in part because it never spent much time asking "*What is the attention area of the brain?*". Nor did the field ask "*Which region implements spatial attention and which implements feature-based attention*". Instead, that research asked the more useful question of "*What is the process by which attention modulates neural activity to promote attentional selection?*" (Desimone and Duncan, 1995; Cohen and Maunsell, 2009). The field has taken a function-first approach instead of an area-first approach.

One important goal should be to develop intellectual tools for understanding how distributed systems can compute. There are many examples in nature of systems that make decisions through distributed consensus. These include bacteria, ants, and so on (Navas-Zuloaga et al., 2022). These systems do not only choose, they also exert control, fail at control, learn, and so on (Eisenreich et al., 2017). It may be that the principles by which these systems perform their computations can help us to understand the principles by which the brain does so. In particular, much of this work emphasizes the idea that agents that are not in direct communication can nonetheless coordinate to implement effective outcomes.

These ideas are distinct from a different set of ideas that can also be useful – those of population-based decisions in neuronal ensembles. Recent work has suggested that the longstanding neuron doctrine – which sees neurons as the core computational unit of the brain – can be replaced with the population doctrine – which sees populations of neurons as the core computational unit (Saxena and Cunningham, 2019; Urai et al., 2022, Ebitz and Hayden, 2021). That idea in turn allows us to understand how cross-areal communication works.

Finally, increased study of how artificial neural networks work, especially with an eye towards understanding general principles of distributed systems, can help us to understand how systems in general can compute in a distributed way.

### Antimodularism and *The Entangled Brain*

Ultimately, my argument is that the existence of brain maps poses a danger to neuroscience because they motivate research practices that reify those maps. In a recent book, *The Entangled Brain*, Pessoa argues that we cannot understand the brain from the standard divide-and-conquer approach (Pessoa, 2022). That is because the brain is a complex entangled system. For example, Pessoa argues that "we need to dissolve boundaries within the brain." Ultimately, Pessoa takes aim at a set of received ideas about the brain that include a modularity in cognitive function and in neuroanatomy (see also Pessoa et al., 2022).

While Pessoa does not directly talk about the perniciousness of areal maps, observations about their malign influence directly tie into themes of *The Entangled Brain*. Indeed, Pessoa specifically argues that we need to downplay brain areas and focus on networks instead, and moreover, we need to be aware that networks reconfigure on fast timescales. Pessoa also argues that brain function is fundamentally emergent, meaning that it arises from the combination of areal properties that are greater than the sum of their parts.

To the extent that *The Entangled Brain* serves as a counterargument to received ideas about the brain, my arguments here provide one explanation for why the field continues to resist Pessoa's ideas, which are good ones. We have these maps, which drive us to see the brain in a certain way. That is not to say that maps are the only reason, or even the most important one – indeed, this is unlikely to be the case. However, my broader theme – that the way we conceptualize the brain influences our study of it – rhymes with several ideas in Pessoa's work.

Pessoa argues that there is a lot of resistance in the field to the idea of the brain as a complex system. This may be true to some extent, but in my experience, there is a lot more sympathy to the idea than one would expect. Instead, I believe the major limiting factor in the adoption of these ideas is the lack of an approachable entry point. That is, lots of people want to take the entangled perspective, they just don't know how to do it. And that's reasonable, because it's a hard problem. Those of us interested in the entangled brain hypothesis need to devote energy to developing those ideas.

Another factor is that those of us who take this view need to acknowledge the many indisputable successes of the modular approach. For example, the identification of motor cortical regions within the brain has led directly to prosthetic devices which show great promise. Likewise, the identification of auditory cortex has led to potential treatments for hearing problems. Even the visual system, which I have used as an example of antimodularity, was studied as a modular system, and the switch in view was driven largely by incoming data inconsistent with the modular theory, not from committed ideologues who adopted an emergentist philosophical position.

Likewise, I think that Pessoa's pessimism about mental categories may be too strong. While I fundamentally agree with him that they are misleading, I would put it differently and say they ought to eventually be *eliminated* (Churchland, 1981). But I am enough of a philosophical pragmatist that I think they may be useful to keep around for a while, even for scientists. I would for example point to the concept of working memory. Its definition has evolved over time as we have grown up and learned more. This approach – adopting a flexibility in definition and a convenient amnesia for past definitions – may be more practical for working scientists than starting from scratch. So perhaps the old definition was problematic, but we have slowly adjusted to newer ones more in keeping with the neuroscience.

### Conclusion

A key concept in Pessoa's article and book are the idea of emergence – how the interaction of multiple parts lead to function that are not observed at the lower levels. This idea has recently gained a great deal of traction in multiple fields. The philosophy of emergence is directly contrary to the animating philosophy behind much of the science of the brain. That is, brain areas, ultimately, seduce us to split instead of lump. We should try something else.